\RequirePackage[2022-09-09]{latexrelease}

\documentclass[amssymb,amsmath,aip,groupedaddress]{revtex4-2}

\usepackage[english]{babel}
\usepackage{graphicx}
\usepackage{color}
\usepackage{bm}
\usepackage{amsmath}
\usepackage{amssymb}

\usepackage[T1,T2A]{fontenc}

\DeclareGraphicsExtensions{.eps,.pdf,.png,.jpg}

\begin{document}

\title{Observation of the diocotron instability in a diode with split cathode}

\author{Y. Bliokh}
\affiliation{Physics Department, Technion, Israel Institute of Technology, Haifa 320003, Israel}
\author{Ya. E. Krasik}
\thanks{Corr.author}
\email{fnkrasik@physics.technion.ac.il}
\affiliation{Physics Department, Technion, Israel Institute of Technology, Haifa 320003, Israel}
\author{J. G. Leopold}
\affiliation{Physics Department, Technion, Israel Institute of Technology, Haifa 320003, Israel}
\author{E. Schamiloglu }
\affiliation{Department of Electrical and Computer Engineering, University of New Mexico, Albuquerque, New Mexico 87131-0001, USA}

\begin{abstract}

Diocotron instability has been observed in the pure electron plasma formed in a split cathode coaxial diode. This plasma consists of electrons, trapped in the longitudinal potential well between the two parts of the cathode. The mathematical model of the electron squeezed state, which allows calculation of the equilibrium plasma density, is presented. The model is applied in a comprehensive analysis of experimental data and the presence of the diocotron instability is unambiguously confirmed.

\end{abstract}

\maketitle

\section{Introduction}

Single-charged non-neutral, pure electron plasmas \cite{Davidson, Oneil-1} are of interest both for basic fundamental \cite{Blaum-2010} plasma physics \cite{Dubin-1999}, including astrophysics \cite{Petri-2008}, atomic physics \cite{Sturm-2013}, and for various practical applications \cite{Katagiri-2021}. Examples of such plasmas are present in the electron cloud in the Malmberg-Penning trap  \cite{Penning-1936,deGrassie-1977}, the squeezed state of electrons \cite{Ignatov-1993, Bliokh-2021}, and electron beams. However, the properties of each of these plasmas are very different. Electrons in the Malmberg-Penning trap are isolated from the environment, their density is quite low, $\leq 10^8$ cm$^{-3}$, and their energy is very low as well (several eV)  \cite{Dubin-1999}. The electron squeezed state is in  permanent contact with the electron source and is characterized by high density, ($\leq 10^{12}$ cm$^{-3}$), and high energy electrons (10s keV)  \cite{Dubinov-2020}. The density and the energy of electron beams can vary over a very broad range. 

In the majority of cases, the geometry of a single-charged plasma is a solid cylinder or an annular column. The diocotron instability \cite{Levi-1965, Davidson}  development results in the radial expansion of the electrons and azimuthal modulation of the electron density. Sometimes this instability is undesirable, but sometimes it is useful. There is some evidence \cite{Riyopoulos-1999} that the diocotron instability can enhance formation of spokes in magnetrons.

In this article, the development of the diocotron instability in a pure electron plasma trapped in the potential well existing between the two parts of a split cathode geometry (see Fig.~\ref{Setup}), is studied experimentally and analytically. Dynamical equilibrium of this plasma is known as a squeezed electron state. Recently, the split cathode was successfully used as a source of electrons in a relativistic magnetron \cite{Krasik-2022}, which explains our interest in this configuration. Diocotron instability in a squeezed state  was observed, but only in numerical simulations \cite{Hramov-2017}.

\section{ EXPERIMENTAL SETUP}

\begin{figure}[tbh]
	\centering \scalebox{0.3}{\includegraphics{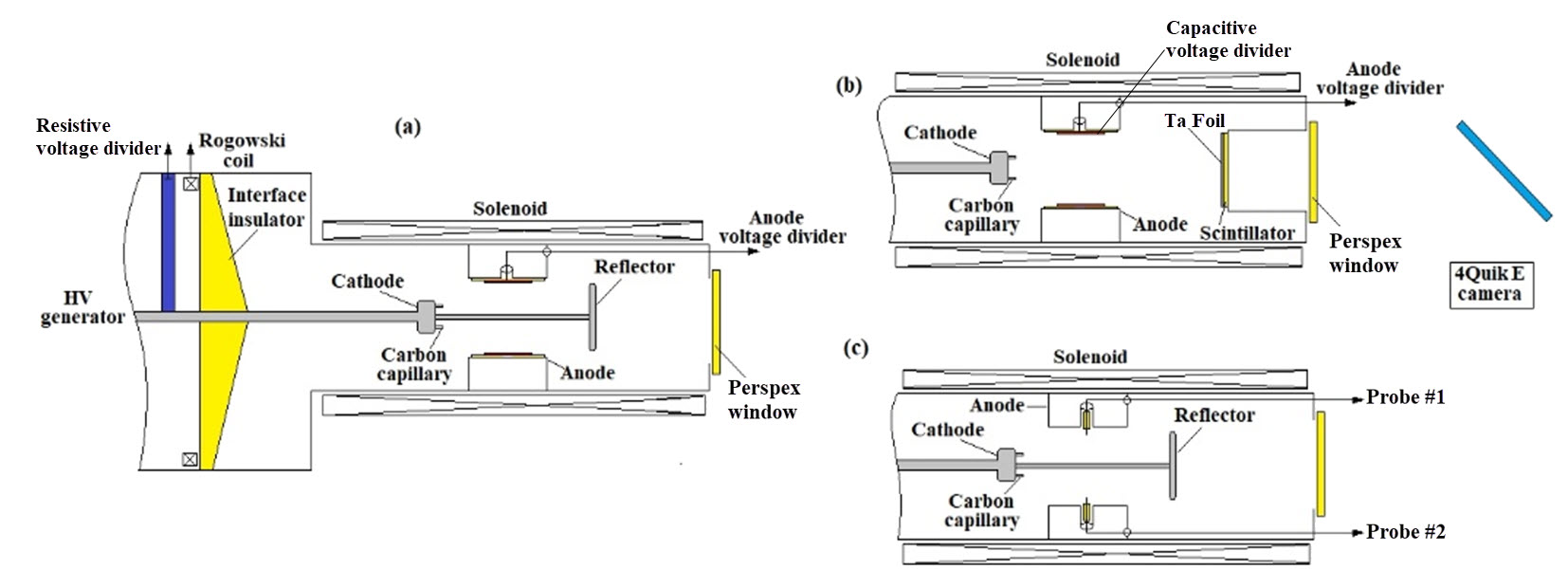}}
	\caption{The experimental setup for (a) a wire grid reflector, (b) to record the beam in the absence of the reflector and rod and (c) with the probes \#1 and \#2. }
	\label{Setup}
\end{figure}

Experiments were carried out using the same experimental setup as that described in detail in \cite{Leopold-2020}
and sketched in Fig.~\ref{Setup}. The experiment was performed in a 124 mm diameter and 600 mm long stainless-steel tube at the center of which a  40 mm diameter, 62 mm long stainless-steel anode insert was placed. An axial magnetic field, $H_0$, of up to 1.2 T  and half-period of 15 ms is produced by a solenoid. The high voltage pulse, formed by a Marx generator ($\sim 200$ kV, $\sim 150$ ns with a resistive load of $\sim 84\Omega$) is applied to 20 carbon-epoxy capillary tubes (each 5 mm long and 1.5 mm/0.75 mm outer/inner diameter) placed along a 16 mm-diameter circle on a 25 mm diameter aluminium cathode. The axial distance between the edge of the capillaries and the anode was 20 mm. The reflector consists of a 2 mm-thick, 40 mm-diameter aluminium disk, and a 6 mm diameter ring at its periphery. The reflector is connected to the cathode by a 6 mm diameter aluminium axial rod. The cathode, reflector and the rod were hard  coated with an 80 $\mu$m-thick Al$_2$O$_3$ ceramic layer to avoid explosive emission plasma formation. The distance between the reflector and the anode  downstream edge was 40 mm. 

The diode voltage and current waveforms were measured using a resistive voltage divider and a self-integrating Rogowski coil, respectively. These probes were placed inside transformer oil tank at the output of the Marx generator. The beam potential was measured using an anode capacitive voltage divider calibrated in shots with a 16 mm diameter rod traversing the anode axis. {This probe was placed at the inner surface of the anode insert.} The diocotron instability appears as rotating azimuthal modulations of the electron density. These modulations were measured by two probes inserted through holes drilled at the longitudinal center of the anode and separated $180^\circ$ from each other. These probes, consist of 80$\mu$m-diameter copper wires, placed inside ceramic capillaries leaving 4 mm bare intended to contact with the electron plasma. The voltage, current, and probe waveforms were acquired by a DSO80604B digitizing oscilloscope.

The temporally-and-spatially-resolved transverse distribution of the electron-beam current density was obtained by fast x-ray imaging of the beam. These experiments were carried out either with a split cathode, with the plate of the reflector replaced by a 40 mm 90\% transparent tungsten wire grid, or without the reflector and the rod. The images were obtained at a distance of 100 mm from the anode where a 2 mm thick fast EJ-200 plastic polyvinyl toluene (PVT) thick-scintillator was placed, attached to the back of a grounded 80 $\mu$m-thick Ta foil. The mean free path of the electrons (electron energy <250 keV) in the Ta foil is less than the foil thickness. However, the mean free path of the x-rays produced by the electrons inside the foil is larger than the foil thickness. The interaction of these x-rays with the scintillator produces a temporally-and-spatially-resolved image, which was recorded using a 4QuikE camera (of 1.2 ns frame duration) synchronized with the high voltage generator. 

\section{Experimental results}

\begin{figure}[tbh]
	\centering \scalebox{0.8}{\includegraphics{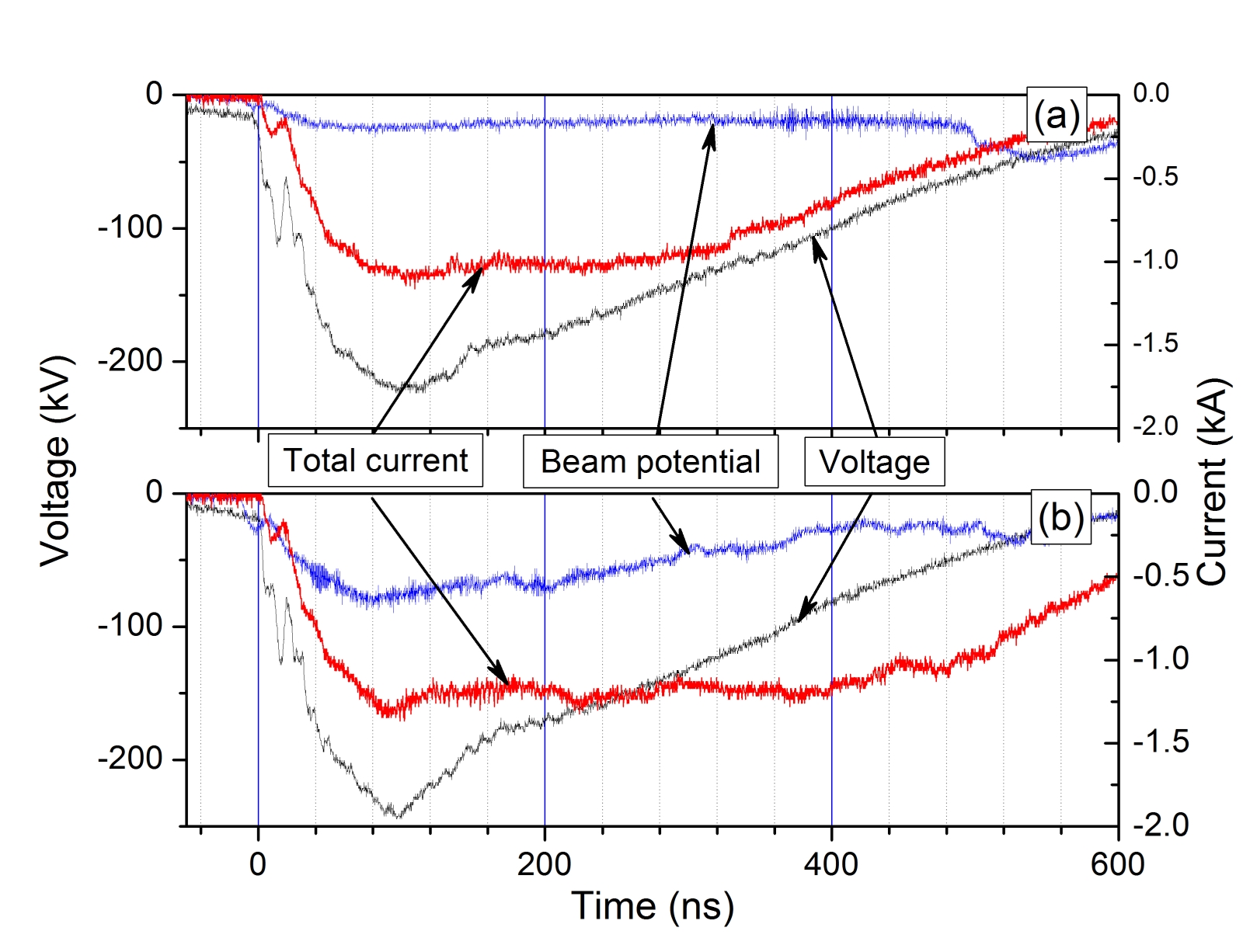}}
	\caption{Waveforms of the voltage, beam current and beam potential (a) without rod and reflector and (b) with the rod but no reflector. The magnetic field here was 6.4 kG }
	\label{Voltage1}
\end{figure}

Time-integrated patterns of the electron beam on the 0.4mm-thick Mylar or 1mm-thick aluminum targets placed at the distance of 10 cm from the anode insert, were obtained in experiments without the reflector and rod. In these experiments, at a magnetic field of 6.4 kG, the electron beam leaves on the targets a $\sim~2.5$ mm thick circular pattern with an average diameter of $\sim18$ mm and without signs of discrete structures. 
In Fig.~\ref{Voltage1}(a), waveforms of the voltage, electron beam current, and the potential of the electron beam are shown for the setup shown in Fig.~\ref{Setup}(a) (no reflector and rod). One can see that the potential of the electron beam is $\sim25$ kV when the beam current is $\sim1$ kA. The $\sim200$ ns long plateau of the current, despite the significant decrease in the amplitude of the applied voltage, is probably the result of the decrease in the effective gap between the boundary of the cathode plasma expanding along the magnetic field upstream from the anode towards it at a velocity of $\sim10^7$ cm/s \cite{Bugaev-1980}, In Fig.~\ref{Voltage1}(b), waveforms of the voltage, electron beam current, and potential when the reflector is absent and only the 6 mm central rod is attached to the cathode, are shown. One can see that the beam potential increases to $\sim75$ kV due to the presence of the rod.

\begin{figure}[tbh]
	\centering \scalebox{0.8}{\includegraphics{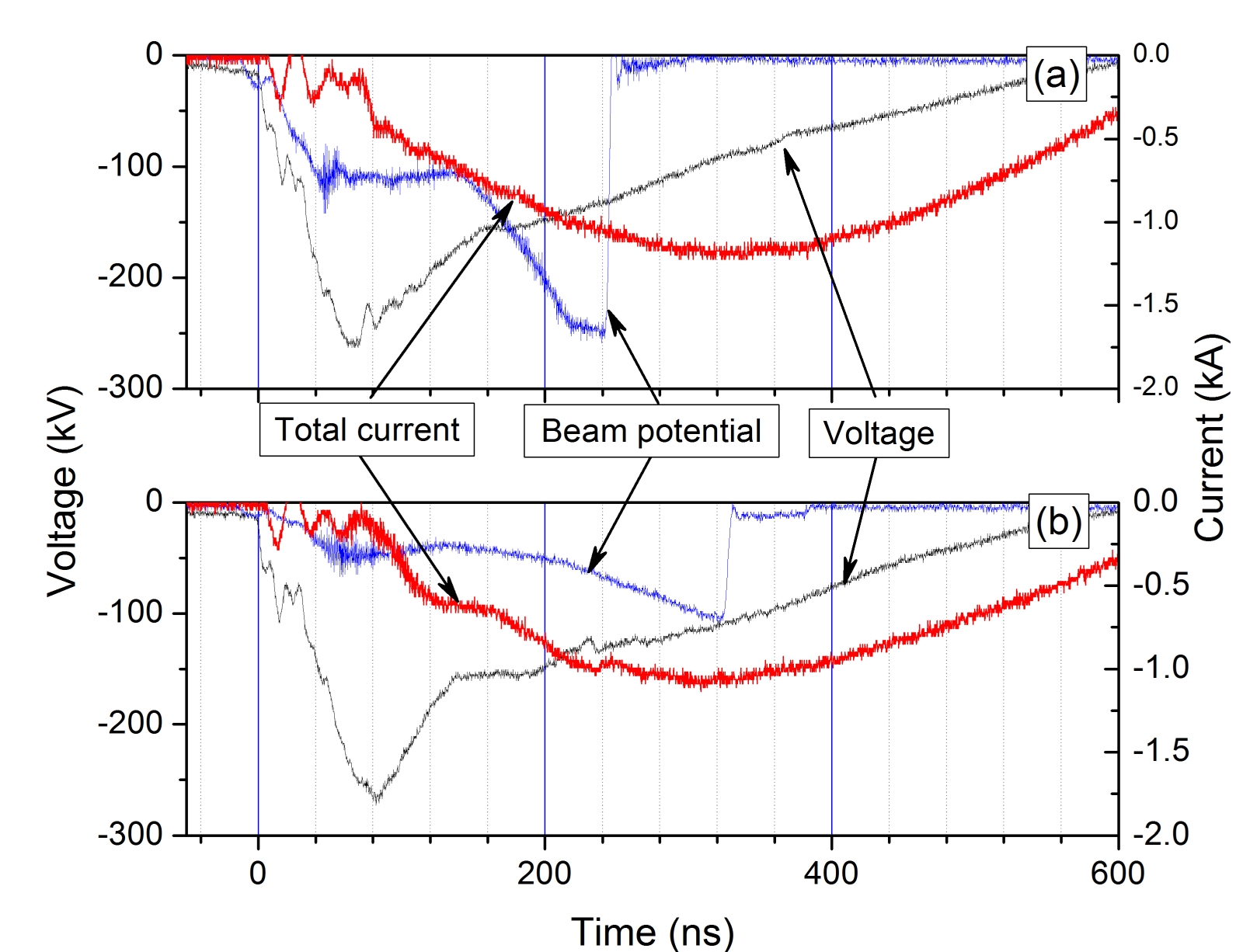}}
	\caption{Waveforms of the voltage, total current and electron beam potential relative to the anode with a split cathode for an applied axial magnetic field of (a) 6.4 kG and (b) 10 kG. }
	\label{Fig3ab}
\end{figure}

In Fig.~\ref{Fig3ab}, the voltage, total current and beam potential in a split cathode configuration for two values of the magnetic field are presented. The first oscillations seen on the current waveform are the result of the displacement current. Next, during $\sim75$ ns, the amplitude of the total current is $\leq150$ A, which is $\sim6.5$ times smaller than when the reflector is absent. However, from $t\approx 80$ ns and on, the total current increases up to $\sim1$ kA at $t\approx 380$ ns. Inspection of the reflector and central rod alumina coating after hundreds of shots showed only a few damage points, indicating that explosive electron emission from these parts are not responsible for the current increase. Thus, the gradual increase in the total current at $t>80$ ns while the voltage decreases can be related to the accumulation of the electrons in the split cathode potential well and the electron space charge radial expansion above cathode radius of 12.5 mm to become an upstream electron flow increasing the total current. For the larger value of the magnetic field in Fig.~\ref{Fig3ab}(b) the current reaches lower values.

The smaller value of the potential for $H_0 = 10$ kG than that obtained for $H_0 = 6.4$ kG (see Fig.~\ref{Fig3ab}) indicates that the space charge accumulated in the potential well between the cathode and reflector increases with increasing magnetic fields. However, one can reasonably consider that this space charge is expanding radially faster for smaller magnetic fields accompanied by a change in the anode voltage divider's coefficient. Since all the quantities, that is, the electron density, its spatial distribution, and the high voltage change together, the signal from the voltage divider presented in Figs.~\ref{Fig3ab}(a) and ~\ref{Fig3ab}(b), cannot be used to determine the value of the electron cloud potential. Nevertheless, a fast increase of the probe signal followed by an abrupt decrease to zero allows one to extract  useful information. This happens when the electron cloud's expansion leads to electron losses to the high voltage electrode of the divider followed by breakdown along the insulator surface. Thus, the breakdown time is the characteristic time of the electron cloud's radial expansion. The results obtained showed that this time grows with increasing magnetic field.

\begin{figure}[tbh]
	\centering \scalebox{0.8}{\includegraphics{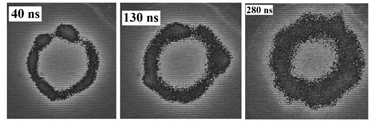}} 
	\caption{Framing images of the luminescence pattern appearing on the plastic scintillator  due to the interaction of the electron beam with a 127-$\mu$m-thick Ta foil placed in front of the scintillator. Frame duration 1.2 ns, magnetic field 8 kG.  }
	\label{Imprint}
\end{figure}

No x-ray electron beam images were obtained in experiments with a split cathode when the aluminum reflector was replaced with a tungsten wire grid. This is because the energy of the electrons near the reflector is strongly reduced and too little of the beam is transmitted. Images were obtained only in the absence of the reflector and registered at different times relative to the onset of the voltage as seen in Fig.~\ref{Imprint} for a magnetic field of $H_0 = 8$ kG. Similar images were obtained for other magnetic fields values as well, independent of whether the rod was present or absent. One can see that the electron beam pattern is not azimuthally uniform and bunching is discernible, typical for a diocotron instability\cite{Davidson}.

\begin{figure}[tbh]
	\centering \scalebox{0.8}{\includegraphics{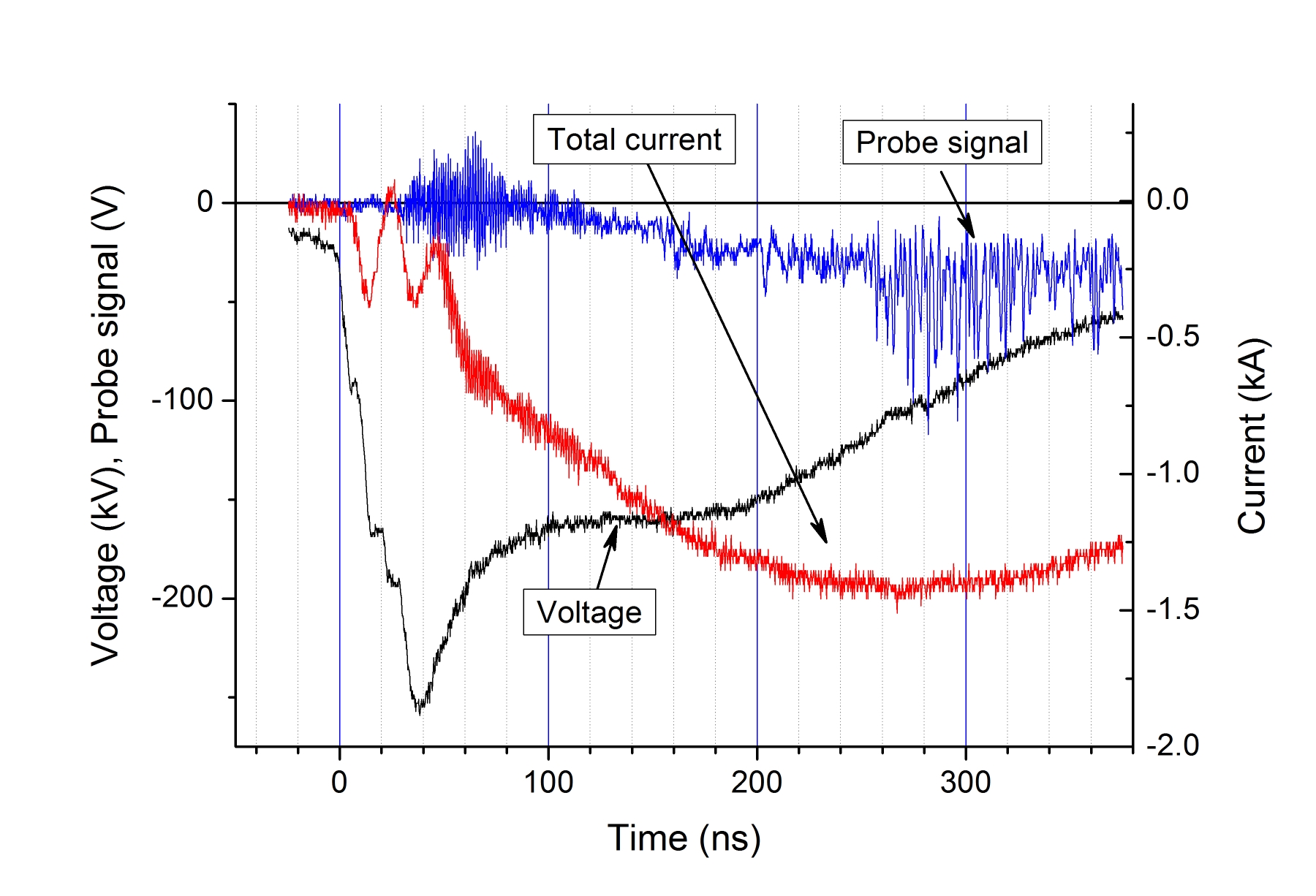}} 
	\caption{The probe signal, voltage, and current waveforms. Split cathode configuration, magnetic field 8 kG. }
	\label{Probes}
\end{figure}

The existence of the diocotron instability in a split cathode geometry is confirmed by high frequency oscillations observed on the probe signals shown, for example, in Fig.~\ref{Probes} together with the voltage and current waveforms. Experiments with probes \#1 and \#2 were carried out in three configurations: (a) the split cathode; (b) no reflector and rod; and (c) no reflector but rod present. In all these cases, high frequency oscillations were observed in the probe signals. Analysis of the waveforms of these probes is in Section IV.

\section {Mathematical model}

\subsection{Steady-State Hollow Electron Cloud}

\begin{figure}[tbh]
	\centering \scalebox{0.3}{\includegraphics{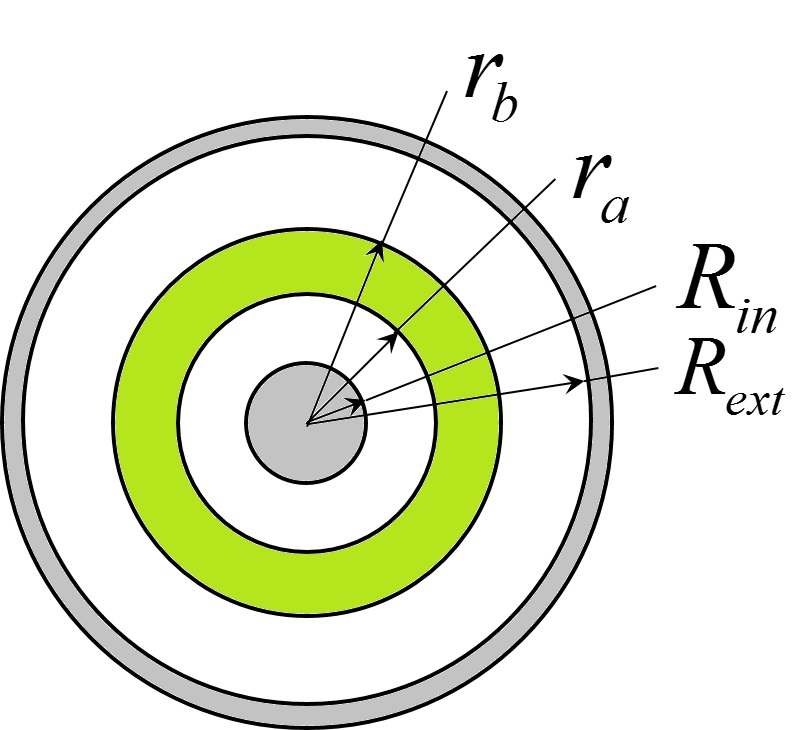}}
	\caption{The electron cloud bounded in the space between cylindrical surfaces of the radii $r_a$ and $r_b$ between the radius $R_{in}$ central rod and the radius $R_{ext}$ of the anode tube. }
	\label{Fig2-1}
\end{figure}

Let us consider an electrons placed in the space between an anode tube of radius $R_{ext}$ and a coaxial central rod of radius $R_{in}$ (Fig~\ref{Fig2-1}). It is assumed that this  cloud is bounded  between cylindrical surfaces of the radii $r_a$ and $r_b$ and has radially-uniform density. In the longitudinal direction the electron motion is restricted by the cathode and reflector  (see Fig.~\ref{Setup}). 
There is a potential difference, $\varphi_0$, between the tube and the rod, and an axial magnetic field $H_0$ which keeps the cloud from expanding radially.  

The  nonrelativistic electron equations of motion are:
\begin{eqnarray}
\label{eq1a}
\frac{dv_r}{dt}=\frac{e}{m}E_r+\omega_Hv_\theta+\frac{1}{r}v_\theta^2,\nonumber\\
\frac{dv_\theta}{dt}=\frac{e}{m}E_\theta-\omega_Hv_r-\frac{1}{r}v_\theta v_r.
\end{eqnarray}   
Here $\omega_H=eH_0/mc$ is the electron gyrofrequency, $v_r$ and $v_\theta$ are electron radial and azimuthal velocities, respectively. 

The steady-state is characterized by zero radial velocity, $v_r=0$, and zero azimuthal electric field, $E_\theta=0$. Then, it follows from Eq.~(\ref{eq1a}), that at steady-state
\begin{equation}
\label{eq2a}
v_\theta(r)^2+r\omega_H v_\theta+\frac{e}{m}rE_r(r)=0
\end{equation}
Introducing the angular frequency $\omega_d(r)=v_\theta/r$ of the rotating electrons, the solution of Eq.~(\ref{eq2a}) can be written as 
\begin{equation}
\label{eq3a}
\omega_d(r)=-\frac{1}{2}\omega_H\pm\sqrt{\frac{1}{4}\omega^2_H-\frac{eE(r)}{mr}},
\end{equation}
The lower frequency corresponds to the rotation with azimuthal velocity $v_\theta$  close to the drift velocity $v_d= cE_r/H_0$, $v_\theta\simeq v_d$ . 

The solution of the Poisson equation in the region $r_a<r<r_b$, the continuity of the potential $\varphi(r)$ and the radial electric field $E_r=-\partial\varphi/\partial r$ at the interfaces $r=r_a$ and $r=r_b$, and the boundary conditions $\varphi(R_{in})=0$ and $\varphi(R_{ext})=\varphi_0$ allow one to determine the potential and electric field  distributions in the entire space $R_{in}\leq r\leq R_{ext}$, as for example in \cite{Levi-1965}. Finally,  the radial electric field allows one to calculate the electrons rotation frequency, $\omega_d(r)$, using Eq.~(\ref{eq3a}). 
Note that the rotating electrons produce additional axial magnetic field so that the gyrofrequency, $\omega_H$, in Eq.~(\ref{eq3a}) depends, strictly speaking, on the radius and should be calculated simultaneously with $\omega_d(r)$. However, for the experimental conditions considered here, this additional magnetic field is small and can be neglected.    

\subsection{Diocotron instability. Dispersion relation}

The diocotron instability, which can develop under certain conditions, breaks the axial symmetry of the described above equilibrium configuration of the electron cloud.  The stability analysis, developed in \cite{Davidson, Levi-1965, Davidson-1998}, is applicable to the system being studied. Note, that a system with a central electrode was considered in \cite{Levi-1965}, but the normalized variables used in this paper are inconvenient when  the electron rotation frequency $\omega_d(r)$ is defined by the external electric field rather than the space-charge field.   

The stability analysis is reduced to the eigenvalue problem for Poisson equation 
\begin{equation}
	\label{Poisson}
	\frac{1}{r}\frac{d}{dr}\left(r\frac{d\phi}{dr}\right)-\frac{\ell^2}{r^2}\phi=-\frac{\ell\, d\omega_p^2(r)/dr}{\omega_Hr[\omega-\ell\omega_d(r)]}\phi
\end{equation}
with boundary conditions $\phi(R_{in})=\phi(R_{ext})=0$
for the potential perturbation of the form $\delta\varphi(r,\theta,t)=\phi(r)\exp(-i\omega t+i\ell\theta)$, $\ell=0,\pm1,\pm2,\ldots$. 

The electron Langmuir frequency $\omega_p(r)$ is defined by the equilibrium distribution of the electron density $n_e(r)$. For the model with a step-function density profile, used in the previous section, the eigenfrequencies $\omega$ are roots of a quadratic equation (dispersion equation), whose derivation and particular form are presented in Appendix A. The general expression for the eigenfrequencies is rather cumbersome. Here we present the solution for a thin electron cloud, the thickness of which $h=r_b-r_a$ is small compared to its mean radius $r_e=(r_b+r_a)/2$, $h\ll r_e$. 

Instead of frequencies $\omega_d(r_{a,b})$, it is convenient to introduce  $\Bar{\omega}_d=[\omega_d(r_a)+\omega_d(r_b)]/2$ and  $\delta\omega_d=[\omega_d(r_a)-\omega_d(r_b)]/2$. The quantity $\Bar{\omega}_d$ is the mean electron rotation frequency, the quantity $\delta\omega_d$ characterizes the difference between the electron rotation frequencies at the inner and outer surfaces of the annular cloud. 

\begin{figure}[tbh]
\centering \scalebox{0.4}{\includegraphics{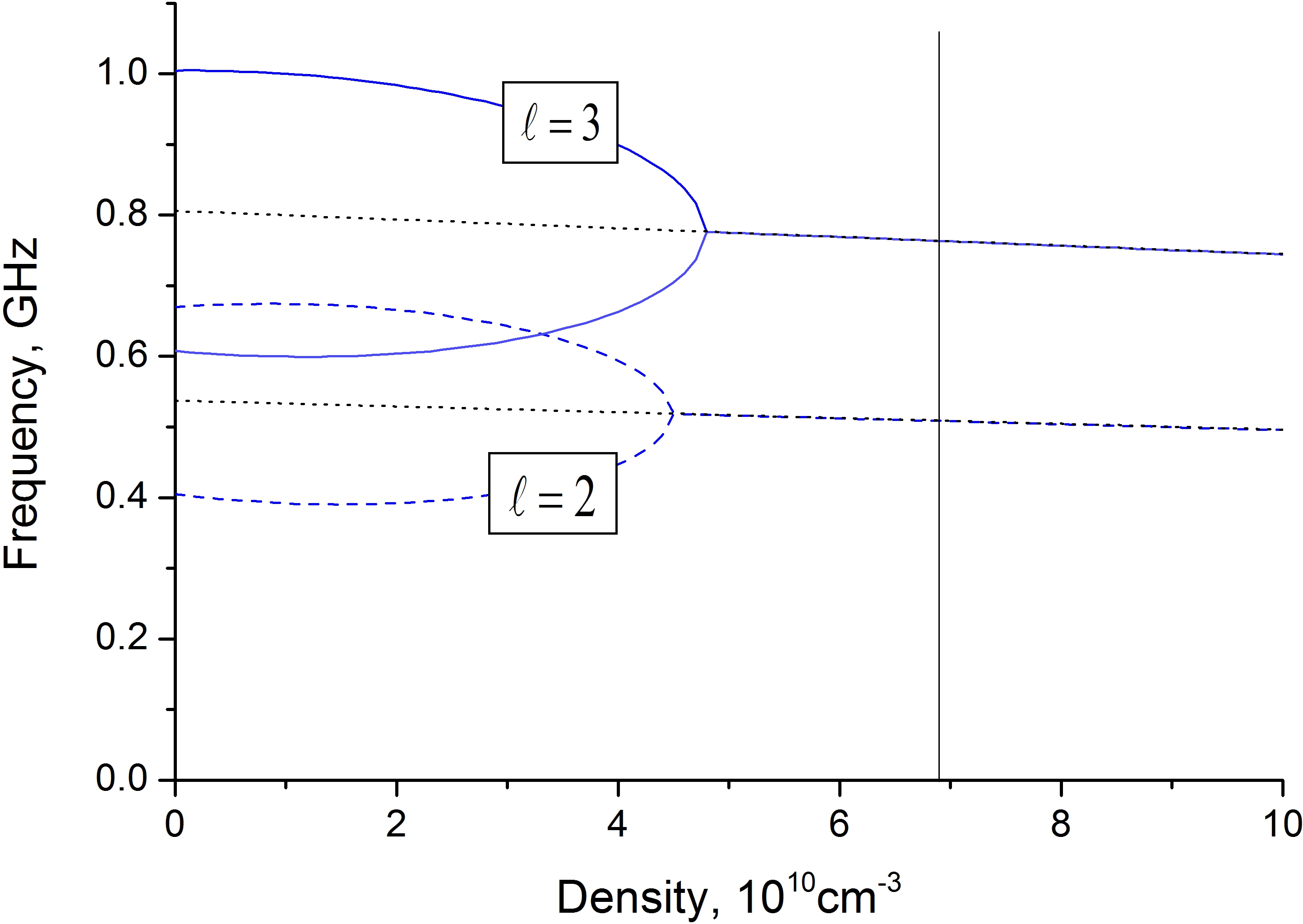}}
\caption{Solution of the dispersion equation (\ref{A7}) for the diocotron modes $\ell=2$ (dashed lines) and $\ell=3$ (solid lines). The rotation frequencies $\ell\Bar{\omega}_d(n_e)$ are shown by dotted lines. Geometrical parameters, magnetic field value, and the voltage between internal and external electrodes correspond to the experimental conditions. The vertical line marks the electron density, calculated within the framework of the squeezed state model.\label{FigDisp}} 
\end{figure}

Assuming that the rather weak conditions $(r_e/R_{ext})^{2\ell}\ll 1$ and $(R_{in}/r_e)^{2\ell}\ll 1$ with $\ell=2,3,..$ are satisfied, the solution of the dispersion equation takes the form (see details in Appendix A)
\begin{equation}
\label{eigenfreq}
\omega\simeq\ell\Bar{\omega}_d\pm\sqrt{\ell\delta\omega_d\left(\ell\delta\omega_d+\frac{\omega_p^2}{\omega_H}\right)-\frac{\omega_p^4\ell h}{2\omega_H^2 r_e} }.
\end{equation}
The electron cloud becomes unstable when the radical expression is negative. 
It follows from Eq.~(\ref{eigenfreq}) that in the instability region $\Re\,\omega\simeq\ell\omega_d(r_e)$, i.e., the azimuthal velocity $\Re\omega r_e/\ell$ of the electron density modulation (electron bunches) is close to the electron rotation velocity $v_\theta\simeq cE_r(r_e)/H_0\simeq\Bar{\omega}_dr_e$. The numerical solution of the dispersion equation (\ref{A7}), presented in Fig.~\ref{FigDisp}, confirms this analysis. Note that the dependence of the mean electron rotation frequency $\Bar{\omega}_d(n_e)$ on the density is weak, $\Bar{\omega}_d(n_e)\simeq\left.\Bar{\omega}_d\right|_{n_e=0}$.

\subsection{The Squeezed Electron State}

The cathode and the reflector in a split cathode form a potential well which traps the emitted electrons while their number increases gradually. To this charge increase, the trapped electrons react by relaxing to a force-free state  (the so-called squeezed state) \cite{Bliokh-2021} characterized by reduced energy and high density  defined by the depth of the potential well restricted radially for very strong magnetic field. For strong but finite magnitude magnetic fields, the external radial electric field can extract electrons from the squeezed state in radial direction. This is the case when the squeezed state can be employed as an electron source with radial emission \cite{Krasik-2022}. 

The steady-state electron density is defined by the balance between the external force that keeps electrons in the potential well and the Coulomb force of the space charge of accumulated electrons in the well. The Coulomb electric field can be calculated  using the Green function method in the same manner as shown in \cite{Smythe,Tien,Rowe} {(for details see the Appendix B)}.

{We now present the annular electron cloud as a set of charged rigid annular rings $r_a\leq r\leq r_b$ of infinitesimal thickness $dz$. 
The expression for the axial component of the electric field $\bar{E}_z(z^\prime,z^{\prime\prime})$ produced in the cross section $z=z^\prime$ by the charged ring located at $z=z^{\prime\prime}$ can be written as 
\begin{equation}
	\label{eq4}
	\bar{E}_z(z^\prime,z^{\prime\prime})=2\pi\sigma {\rm sign}(z^\prime-z^{\prime\prime})G(|z^\prime-z^{\prime\prime}|),
\end{equation} 
where $\sigma=en_edz$ is the surface charge density, ${\rm sign}(z^\prime-z^{\prime\prime})=1\, {\rm or}\, -1$ for $z^\prime>z^{\prime\prime}$ and $z^\prime<z^{\prime\prime}$, respectively.}

{It was shown in \cite{Tien, Rowe, Morey} that the function $G(|z^\prime-z^{\prime\prime}|)$ for a solid cylindrical electron cloud can be approximated by the exponential
\begin{equation}
	\label{eq20a}
	G(|z^\prime-z^{\prime\prime}|)\simeq \exp(-|z^\prime-z^{\prime\prime}|/\ell_C),
\end{equation}
where the characteristic scale of the Coulomb field  $\ell_C\simeq r_b/2$. 
An annular cloud is characterized by additional of geometrical parameters. Nevertheless, the exponential approximation Eq.~(\ref{eq20a}) is still applicable (see Fig.~\ref{FigA1}). }
{The dimensions of the conductors and the electron cloud in the experiment described in Section III, are $r_a=0.7$cm, $r_b=0.9$cm, $R_{in}=0.3$cm, and $R_{ext}=2.0$ cm.  The characteristic scale of Coulomb field for these parameters is $\ell_C\simeq 0.56$cm. }

{The equation of motion of the charged rings, comprising the electron cloud in the external and the self-consistent space charge electric fields can be written as \cite{Bliokh-2021}
\begin{equation}
	\label{eq21a}
	\frac{d^2z}{dt^2}=-\frac{e}{m}\frac{\partial\varphi}{\partial z}+2\pi\frac{e^2}{m}\intop_0^Ldz^\prime n_e(z^\prime){\rm sign}(z-z^\prime) G(|z-z^\prime|).
\end{equation}   
Here $\varphi$ is the external potential and $L$ is the system length. 
Assuming that the characteristic spatial scale of the variation  of the external potential $\varphi$ and the electron density $n_e$ is large  compared to $\ell_C$, one can evaluate the integral in Eq.~(\ref{eq21a}) as
\begin{equation}
	\label{eq22a}
	\intop_0^Ldz^\prime n_e(z^\prime){\rm sign}(z-z^\prime) G(|z-z^\prime|)\simeq\intop_0^Ldz^\prime n_e(z^\prime){\rm sign}(z-z^\prime) e^{(|z-z^\prime|)/\ell_C}\simeq 2\frac{dn_e(z)}{dz}\ell_C^2.
\end{equation} 
It follows from Eq.~(\ref{eq22a}) that the steady-state solution exists for 
\begin{equation}
	\label{eq23a}
	n_e(z)=\frac{\varphi(z)}{4\pi e\ell_C^2}\simeq 5.7\cdot 10^8 \frac{\varphi(z)[kV]}{\ell_C^2[cm]} [cm^{-3}].
\end{equation}
Note, that  $\varphi(z)$ is the potential profile along the electron propagation path, i.e.,  $\varphi(z)=\varphi(z,r)_{r={\rm const}}$.}

{The depth  of the potential well, $\varphi_0$, where the squeezed electron state (cloud) forms, is defined by the potential profile along the line $r=r_e\simeq (r_b+r_a)/2$ of the electron trajectory. In the configuration with a central rod, $\varphi_0=U_0(|\ln\rho_e/\ln\rho_{in}|-1)\simeq 0.5U_0$, where $\rho_{e}=r_e/R_{ext}$, $\rho_{in}=R_{in}/R_{ext}$ and $U_0$ is the potential difference between the central electrode and the external tube (anode). In the experiments (see Section V for details) $U_0=75$kV so that $\varphi_0\simeq 37.5$kV. Using  $\ell_C\simeq 0.56$cm and $\varphi_0\simeq 37.5$kv, in Eq.~(\ref{eq23}), the electron density can be estimated as $n_e\simeq 6.9\cdot 10^{10}$cm$^{-3}$.}

\section{Analysis of the Experimental Results} 

\begin{figure}[tbh]
	\centering \scalebox{0.6}{\includegraphics{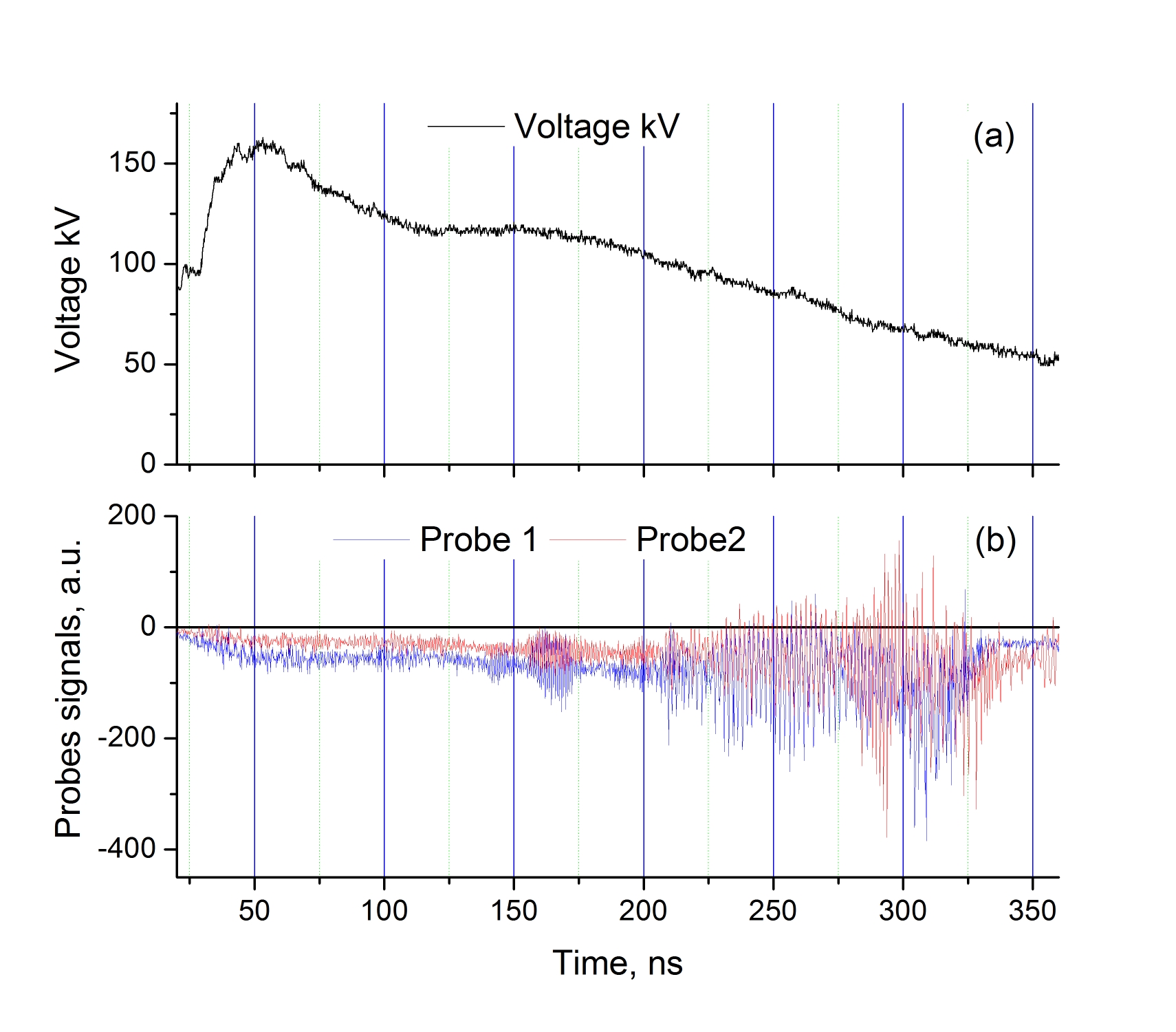}}
	\caption{(a) Potential difference between anode and the rod electrode and (b) two probe signals.}
	\label{Fig2}
\end{figure}

As mentioned in Section II, two probes were placed at two diametrically opposite locations along the anode circumference. Typical waveforms of the probe signals and the potential difference between the anode and  central rod are presented in Fig.~\ref{Fig2}.

\subsection{Processing the Probe Signals }

\begin{figure}[tbh]
	\centering \scalebox{0.6}{\includegraphics{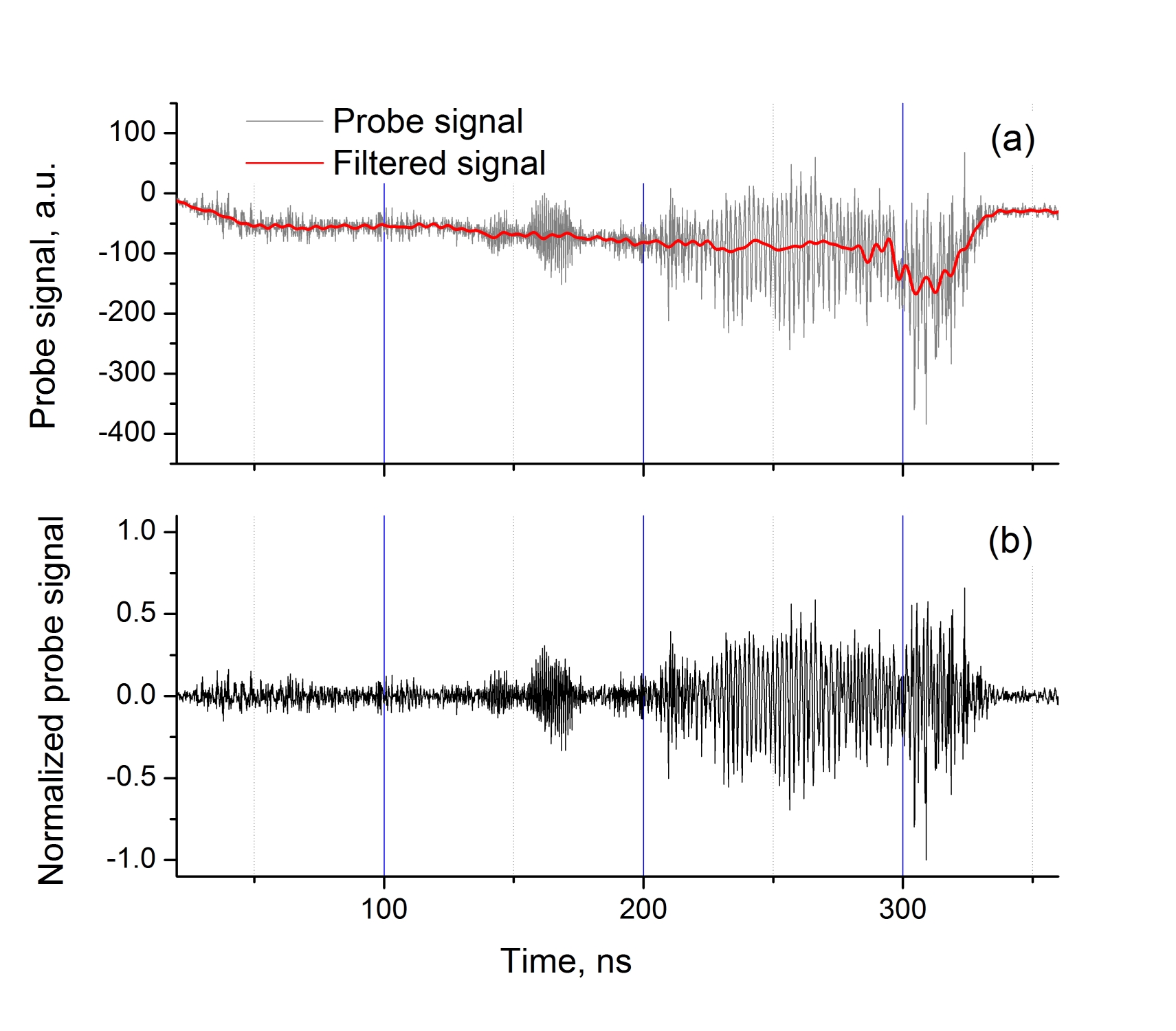}}
	\caption{(a) Raw probe signal (gray) and filtered base line (red).  (b) Normalized fast oscillating component.}
	\label{Fig3}
\end{figure}

First,  the fast oscillating component of the probe signals was separated from the slowly varying basis as illustrates in Fig.~\ref{Fig3}.

\begin{figure}[tbh]
	\centering \scalebox{0.8}{\includegraphics{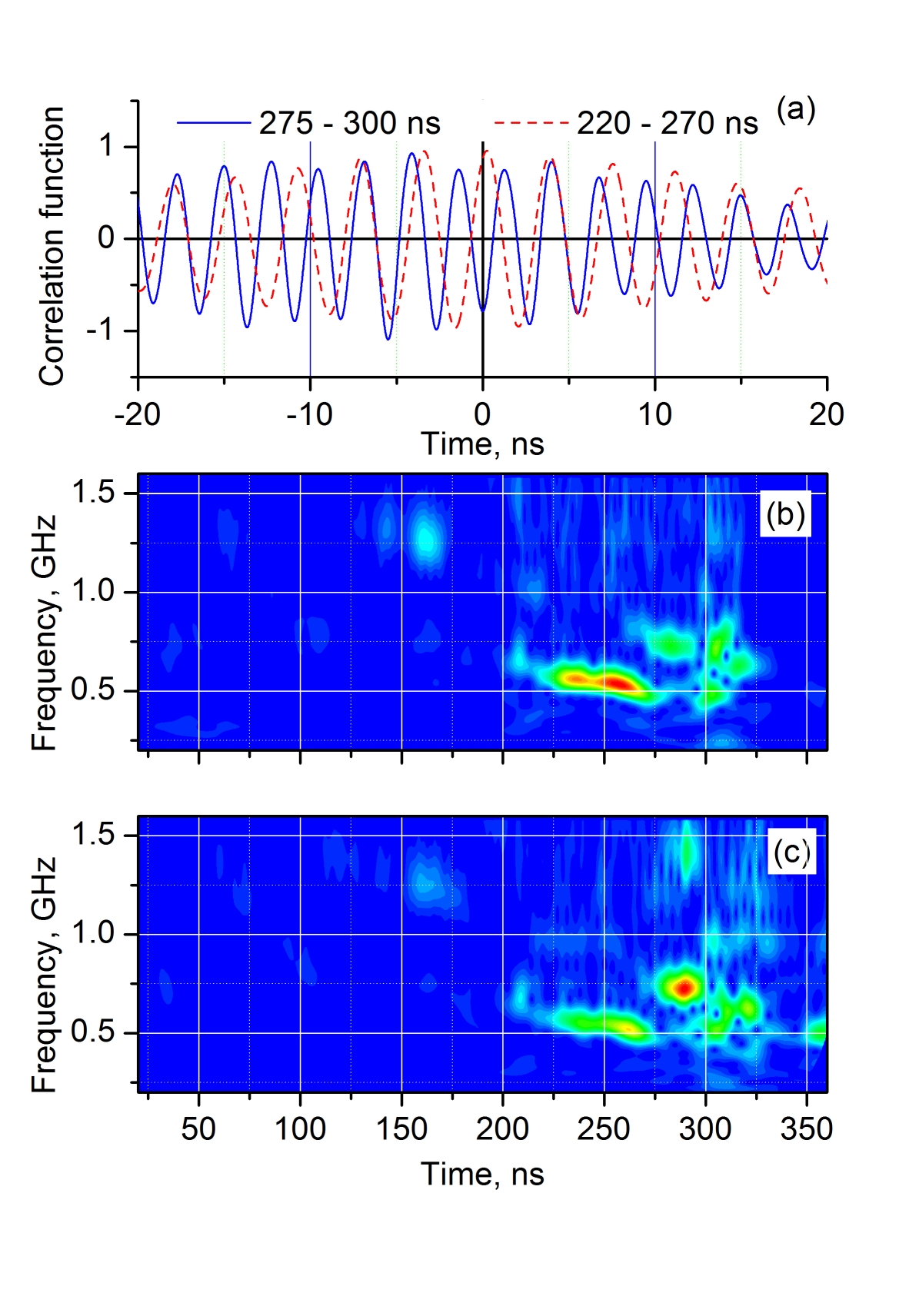}}
	\caption{(a) Correlation functions of the signals within the time intervals 275 - 300 ns and 220 - 270 ns; (b) and (c) Spectrograms of the signals registered by the two probes. }
	\label{Fig4}
\end{figure}

Next, a wavelet analysis of the normalized filtered signals is performed. The resulting spectrograms of two probe signals are seen in Fig.~\ref{Fig4}.
One can see a sharp change in the spectra in the time interval 275-300 ns. It seems reasonable to analyze the correlation between the signals from two probes in these two time domains, 220-270 ns and 275-300 ns, separately. The resulting correlation functions are shown in Fig.~\ref{Fig4}(a). These will be discussed in Section V.B.

\subsection{Data Analysis}

\begin{figure}[tbh]
\centering \scalebox{0.6}{\includegraphics{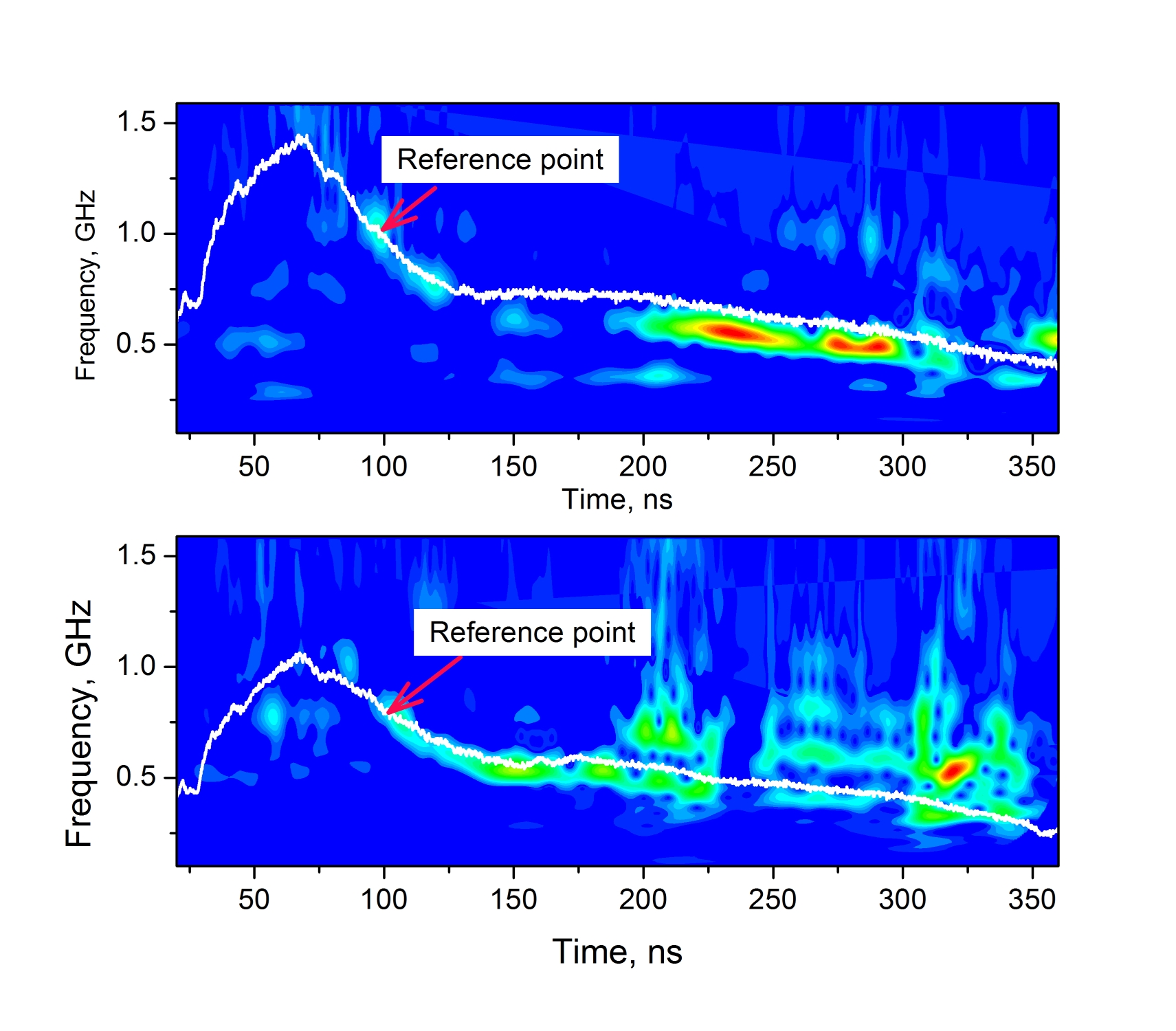}}
	\caption{Two different spectrograms and the corresponding scales voltages. The voltage waveforms $U(t)$ are scaled so that they coincide with the dominant frequency  at some point marked as the reference point. }
	\label{Fig5}
\end{figure}

Periodic probe signals appear as the result of rotating electron {density modulation} produced by the diocotron instability.  The rotation frequency of the density modulation, as shown in Eq.~(\ref{eigenfreq}), is close to the electron rotation frequency, defined by the drift velocity $v_d=cE/H$ in the crossed radial electric, $E$, and axial magnetic, $H$, fields. In order to support this statement, we check that the temporal variations of the voltage and the spectrum correlate. To do this, let us superimpose a spectrogram and an appropriately scaled voltage waveform  as shown in Fig.~\ref{Fig5}.  
It can be clearly seen that the time dependence of the voltage waveform coincidences with the variation in the spectrum over a long period of time. This indicates that the diocotron instability can indeed be responsible for the observed oscillating signal behavior. Note, that similar methods to detect the diocotron instability using azimuthally separated probes or a segmented anode cylinder, was described in \cite{Rosental-1987, Bettega-2008}.

\begin{figure}[tbh]
	\centering \scalebox{0.8}{\includegraphics{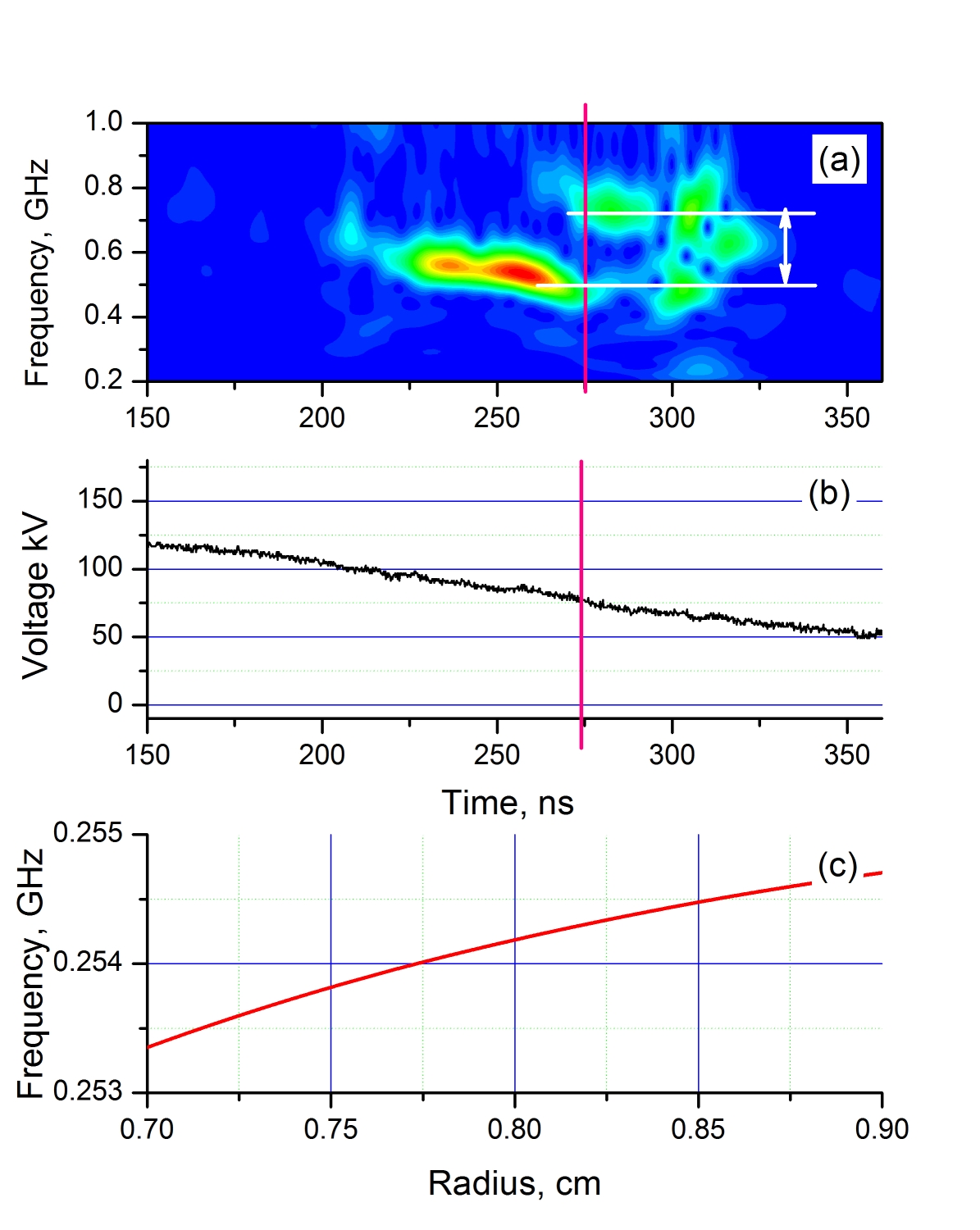}}
	\caption{Experimental determination of the rotation frequency of the bunches. (a) Spectrogram. Vertical red line marks the point in time when the frequency jump occurs. The horizontal white lines mark the values of the frequencies.  (b) The voltage waveform. Vertical red line marks the voltage value at the time the frequency jump occurs. (c) The rotation frequency, calculated using the theoretical model, described in Section~IV. }
	\label{Fig6}
\end{figure}

The raw data used in Fig.~\ref{Fig4} demonstrates an additional property, which can also be considered as  manifestation of the diocotron instability. The correlation functions, presented in Fig.~\ref{Fig4}(a), show that the oscillating signals are in-phase in the time domain $220<t<270$ ns, and out of phase in the time domain $275<t<300$ ns. {This can be explained by the fast transformation of the azimuthally rotating electron column  from even to odd number of the azimuthal high-density bunches of electrons.} If this assumption is correct, then the difference between the frequencies in the time domains is the \textit{experimental} determination of the rotation frequencies of the bunches, as shown in Fig.~\ref{Fig6}. 

Calculation of the rotation frequency, $f(r)=\omega(r)/2\pi$, was performed using the model described in Section~IV. The frequency jump occurs when the voltage is  $\simeq 75$kV. Following Section~IV, the corresponding electron density of the squeezed state at this moment is  $\simeq 6.9\cdot 10^{10}$cm$^{-3}$. The calculated dependence $f(r)$ for these values of the voltage and the electron density for the experimental system dimensions is shown in Fig.~\ref{Fig6}c. 
The calculated frequency value agrees well with the defined experimentally difference $\Delta f\simeq 0.25$GHz  between frequencies, shown in Fig.~\ref{Fig6}(a). The minor frequency from the considered is nearly twice as large as the rotation frequency, while the higher frequency is approximately three times larger. It means, that the lower frequency corresponds to the electron cloud configuration with two (even!) {azimuthal bunches}, the higher frequency corresponds to configuration with three (odd!) {bunches}. The latter agrees with the result of correlation analysis, presented in Fig.~\ref{Fig4}(a). 

These results -- number of azimuthal bunches and modulation frequencies -- agree perfectly with numerical solution of the dispersion equation, presented in Fig.~\ref{FigDisp} in Section IV B.  

\section{Conclusions}

The split cathode is a promising source of electrons for high power microwave devices. Electrons, emitted by an annular cathode, are trapped in the longitudinal potential well forming electron cloud, the radial expansion of which is limited by a strong axial magnetic field. First experimental studies of a relativistic magnetron with a split cathode confirmed that the absence of the cathode explosive emission plasma inside the anode interaction space prevents microwave pulse shortening \cite{Krasik-2022}.
	
The azimuthal symmetry of an electron cloud in a magnetron can be violated by two processes. First, the interaction of the electrons with electromagnetic fields of the magnetron cavity eigenmodes. Second, by the diocotron instability, which for the split cathode geometry, has not been studied so far. In the present article, for the split cathode fed smooth bore anode, it has been shown, by a comprehensive theoretical analysis of the experimental data, that the diocotron instability indeed develops. Thus, we expect that for a corresponding magnetron, the diocotron instability and the magnetron eigenmodes could act simultaneously and influence each other. Matching between the diocotron and magnetron eigenmodes can improve the magnetron characteristics.

\section*{Acknowledgments} 

The authors are grateful to two anonymous referees for valuable comments that have lead to a substantial improvement in the manuscript.
At the Technion, this work was supported by Technion (Grant No. 2029541) and ONRG (Grant No. N62909-21-1-2006).  We are thankful for the assistance in experiments Y. Cao, S. Gleizer and E. Flyat. At the University of New Mexico, this work was supported by ONR (Grant No. N00014-19-1-2155).

\section*{Data availability}

The data that support the findings of this study are available
from the corresponding author upon reasonable request


\section*{Appendix A}

In this Section we apply the stability analysis, developed in \cite{Levi-1965, Davidson-1998} to the system with step-like profile of the electron cloud, considered in Section IV A.
 
Normalizing all frequencies to $c/R_{ext}$ ($\Omega=\omega R_{ext}/c$, $\Omega_p=\omega_p R_{ext}/c$, etc.) and radii to $R_{ext}$ ($\rho=r/R_{ext}$, $\rho_a=r_a/R_{ext}$, etc.), let us write Eq.~(\ref{Poisson}) in the dimensionless form: 
\begin{equation}
	\label{A1}
	\frac{d}{d\rho}\left(\rho\frac{d\phi}{d\rho}\right)-\frac{\ell^2}{\rho}\phi=-\frac{\ell\, d\Omega_p^2(\rho)/d\rho}{\Omega_H[\Omega-\ell\Omega_d(\rho)]}\phi.
\end{equation} 
For the step-like profile of the electron density, $n_e(\rho)=n_0$ when $\rho\in(\rho_a,\rho_b)$ and $n_e=0$ elsewhere, 
\begin{equation}
	\label{A2}
	d\Omega_p/d\rho=\Omega_{p0}[\delta(\rho-\rho_a)-\delta(\rho-\rho_b)],
\end{equation}
where $\delta(x)$ is the Dirac $\delta$-function.

General solution of Eq.~(\ref{A1}) is
\begin{equation}
	\label{A3}
	\phi=\alpha\rho^\ell+\beta\rho^{-\ell},
\end{equation}
where $\alpha$ and $\beta$ are arbitrary constants. Using the boundary conditions $\phi(\rho_{in})=\phi(1)=0$ and taking into account Eq.~(\ref{A2}), the solution of Eq.~(\ref{A1}) can be written as
\begin{eqnarray}
\phi\equiv\phi_{\rm I}=A\left(\rho^\ell-\rho^{-\ell}\rho_{in}^{2\ell}\right),\hspace{5mm}\rho\in(\rho_{in},\rho_a),\nonumber\\
\phi\equiv\phi_{\rm II}=C\rho^\ell+D\rho^{-\ell},\hspace{5mm}\rho\in(\rho_a,\rho_b),\label{A4}\\
\phi\equiv\phi_{\rm III}=B\left(\rho^\ell-\rho^{-\ell}\right),\hspace{5mm}\rho\in(\rho_b,1),\nonumber 
\end{eqnarray} 
Function $\phi$ is continuous everywhere over the interval $(\rho_{in},1)$
\begin{eqnarray}
\phi_{\rm I}(\rho_a)=\phi_{\rm II}(\rho_a),\nonumber\\
\phi_{\rm III}(\rho_b)=\phi_{\rm III}(\rho_b),\label{A5}	
\end{eqnarray}
and its derivative $\phi^\prime =d\phi/d\rho$ experiences stepwise variations at the interfaces between the regions:
\begin{eqnarray}
\phi^\prime_{\rm II}(\rho_a)-\phi^\prime_{\rm I}(\rho_a)=-\frac{\ell\Omega_p^2\phi(\rho_a)}{\rho_a\Omega_h\left[\Omega-\ell\Omega_d(\rho_a)\right]},\nonumber\\
\phi^\prime_{\rm III}(\rho_b)-\phi^\prime_{\rm II}(\rho_b)=\frac{\ell\Omega_p^2\phi(\rho_b)}{\rho_b\Omega_h\left[\Omega-\ell\Omega_d(\rho_b)\right]}.\label{A6}	
\end{eqnarray} 

The dispersion equation appears as a compatibility condition for the system (\ref{A5})--(\ref{A6}). Introducing $\Bar{\Omega}_d=[\Omega_d(\rho_a)+\Omega_d(\rho_b)]/2$, $\delta \Omega=[\Omega_d(\rho_a)-\Omega_d(\rho_b)]/2$, and $w=\Omega-\ell\Bar{\Omega}_d$, 
one can write the dispersion equation in the following form:
\begin{equation}
	\label{A7}
	\alpha w^2+\beta w+\gamma=0.
\end{equation}
Here 
\[\alpha=\Delta_-(1-g_ag_b)-\Delta_+(g_a-g_b),\hspace{3mm} \beta=\mu\Delta_-(g_a+g_b),\]
\[\gamma=\Delta_-\mu^2+\ell\mu\delta\Omega[2\Delta_+-\Delta_-(g_a-g_b)]- \alpha \ell^2\delta\Omega^2,\hspace{3mm} \mu=\Omega_p^2/\Omega_H,\]
\[\Delta_-=\rho_a^\ell\rho_b^{-\ell}-\rho_a^{-\ell}\rho_b^\ell,\hspace{3mm} \Delta_+=\rho_a^\ell\rho_b^{-\ell}+\rho_a^{-\ell}\rho_b^\ell,   \]
\[g_a=[1+(\rho_{in}/\rho_a)^{2\ell}]/[1-(\rho_{in}/\rho_a)^{2\ell}],\hspace{3mm} g_b=(1+\rho_b^{2\ell})/(1-\rho_b^{2\ell}). \]

\begin{figure}[tbh]
	\centering \scalebox{0.4}{\includegraphics{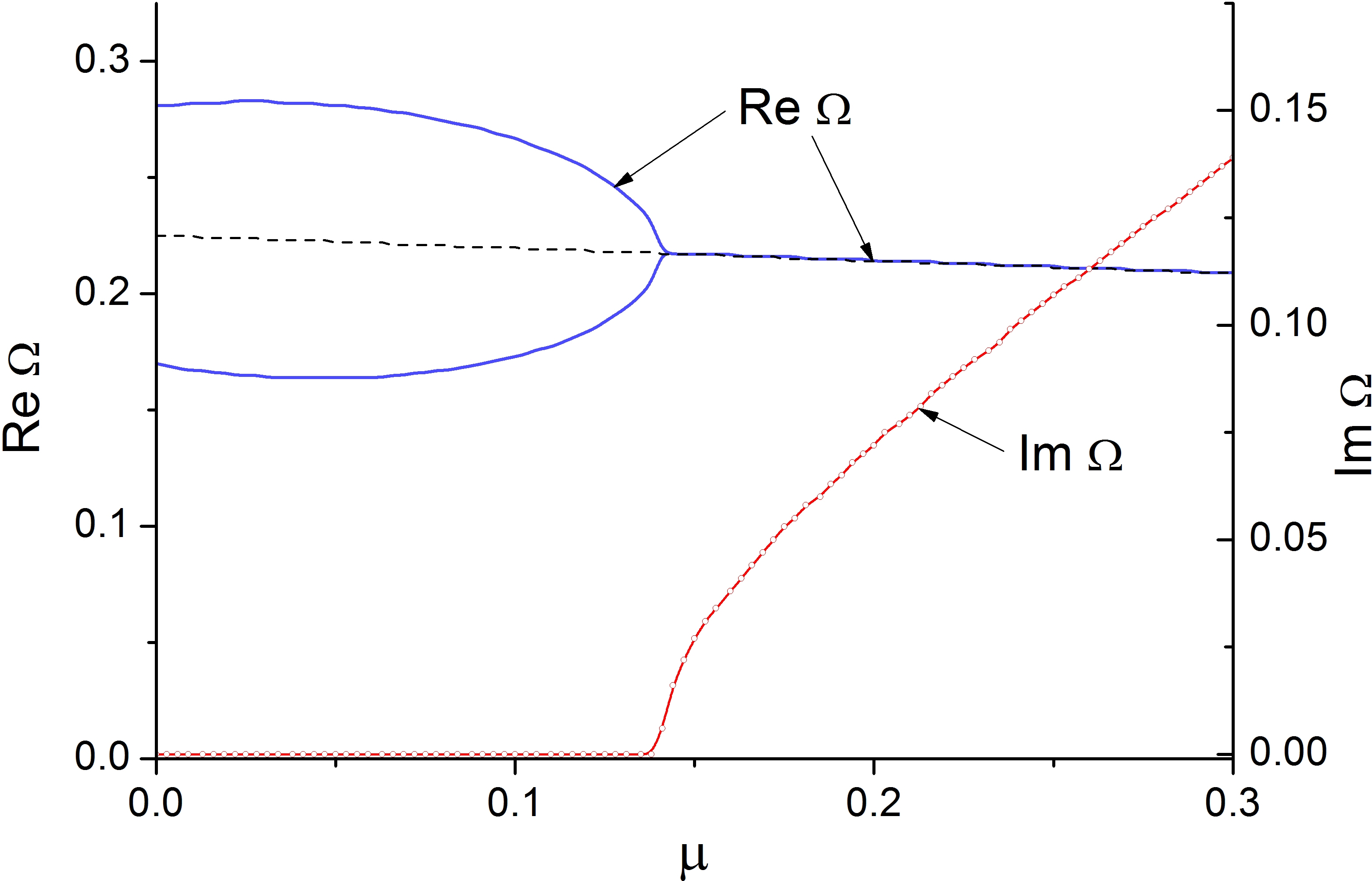}}
	\caption{The solution $\Omega$ of the dispersion equation (\ref{A7}) as function of the parameter $\mu$. The dashed line is the value of $\ell\Bar{\Omega}_d$ vs $\mu$. The geometrical parameters are the same as in the experiment. In the instability region ${\rm Re}\,\Omega=\ell\Bar{\Omega}_d$ }
	\label{FigA}
\end{figure}

For the qualitative analysis of the dispersion equation, it is convenient to consider a thin annular cloud, assuming that $\rho_b-\rho_a=d\ll\rho_a+\rho_b=2\rho_e$, where $d$ is the cloud thickness and $\rho_e$ is the cloud mean radius. Assuming also that $(\rho_{in}/\rho_e)^{2\ell}\ll1$ and $\rho_e^{2\ell}\ll1$, where $\ell=2,3,\ldots$, the dispersion equation (\ref{A7}) can be simplified to
\begin{equation}
	\label{A8}
	w^2=\ell\delta \Omega(\ell\delta \Omega+\mu)-\ell\mu^2\delta/2\rho_e.
\end{equation}
The diocotron instability develops when the right-hand side of Eq.~(\ref{A8}) is negative. For this case 
\begin{equation}
	\label{A9}
	{\rm Re}\,\Omega=\ell\Bar{\Omega}_d.
\end{equation}
This result is confirmed by the solution of Eq.~(\ref{A7}), presented in Fig.~\ref{FigA}. One can see that the dependence of the mean electron rotating frequency $\Bar{\Omega}_d$ on the electron density is weak, so that its value can be calculated for  a ``cold'' system, when the space charge field of the electron cloud is small compared to the external electric field.  Eq.~(\ref{A9}) means that the azimuthal modulation of the electron density rotates practically with the same velocity as the electrons themselves.

\section*{Appendix B}

\subsection{The Space Charge Electric Field}

Consider the electron cloud occupying the  annular region $r_a\leq r\leq r_b$ arranged coaxially in a conducting tube of radius $R_{ext}$, as shown in  Fig.~\ref{Fig2-1}.  In general, the conducting tube can also contain a central rod.  {Let us find the potential $\varphi({\bf r})$ induced by an annular disk $r_a\leq r\leq r_b$ of infinitesimal thickness $dz$ in the $z$ direction.}

The electrostatic potential, produced by an axially-symmetrical charged layer, placed at $z=0$,  in the half-space $z\geq 0$, can be written as
\begin{equation}
	\label{eq1}
	\varphi({\bf r})=\sum_n A_ne^{-\mu_nz}F_n(r),
\end{equation}    
where $F_n(r)$ are the eigenfunctions of the Laplace equation
\[\frac{1}{r}\frac{d}{dr}\left(r\frac{dF_n}{dr}\right)+\mu_n F_n=0    \]
with zero boundary conditions either in the ring area $R_{in}\leq r\leq R_{ext}$ or in the circular area $0 \leq r\leq R_{ext}$ in the absence of the central conducting rod.  Here $\mu_n$ are the eigenvalues. 
Because of symmetry, on the entire $z = 0$ plane,  $E_z|_{z=0}=-(\partial\varphi/\partial z)_{z=0}=0$, 
except on the charged layer itself at $r\in (r_a,r_b)$.  Differentiating Eq.~(\ref{eq1}) and setting $z = 0$, gives
\begin{equation}
	\label{eq2}
	\left(\partial \varphi/\partial z\right)_{z=0}=-\sum_n A_n\mu_nF_n(r).
\end{equation} 
The coefficients $A_n$ of the expansion Eq.~(\ref{eq1}) can be calculated as
\begin{equation}
	\label{eq3}
	A_n= -\frac{1}{\mu_n\|F_n\|^2}\intop_{r_a}^{r_b}\left(\frac{\partial \varphi}{\partial z}\right)_{z=0}rF_n(r)dr,
\end{equation}
where $\|F_n\|=\sqrt{\intop_{R_{in}}^{R_{ext}}F_n^2(r)rdr}$ is the norm of the eigenfunction.

The electric field ${\bf E}$ at the surface of charged layer is directed along the $z$-axis and is equal to $2\pi\sigma$, where $\sigma$ is the layer surface charge density. Thus, 
\[E_z|_{z=0}=-\left(\partial\varphi/\partial z\right)_{z=0}=2\pi\sigma.  \] 
Assuming, that the layer is charged homogeneously, one can rewrite Eq.~(\ref{eq3}) as
\[ A_n= \frac{2\pi\sigma}{\mu_n\|F_n\|^2}\intop_{r_a}^{r_b}rF_n(r)dr.  \]  
Expression (\ref{eq1}) acquires the form
\begin{equation}
	\label{eq4a}
	\varphi(r,z)=2\pi\sigma\sum_n \mu_n^{-1}\|F_n\|^{-2}e^{-\mu_nz}F_n(r)\intop_{r_a}^{r_b}rF_n(r)dr.
\end{equation}
Accordingly, the axial electric field reads as
\begin{equation}
	\label{eq5}
	E_z(r,z)={\rm sign}(z)2\pi\sigma\sum_n \|F_n\|^{-2}e^{-\mu_n|z|}F_n(r)\intop_{r_a}^{r_b}rF_n(r)dr,
\end{equation}
where ${\rm sign}(z)=1\, {\rm or}\, -1$ for $z>0$ and $z<0$, respectively.

Let the radial thickness $d$ of the electron cloud be small compared to the external cylinder radius $R_{ext}$. For $r_a=0$ and no central rod, the 
cloud cross section is a solid disk.  Consider the disk radius $r_b$ be small, $r_b \ll R_{ext}$. Then, assuming that the electron density $n_e$ is homogeneous in the transverse plane, the radial dependence of the electric field, $E_z$, which governs the axial motion of the electrons comprising the layer $dz$, can be neglected and the problem reduces to the one-dimensional. 

Following Tien \textit{et al.}\cite{Tien}, let us replace $dz$-thick slices of the electron cloud  by charged rings (or disks) with the same total charge so that $\sigma=en_edz$. In the model of rigid slices, the acting electric field $E_z(r,z)$ should be replaced by its average over the slice surface area 
\[ \bar{E}_z(z)=\frac{2}{r_b^2-r_a^2}\intop_{r_a}^{r_b} E_z(r,z)rdr.\]
Thus,
\begin{equation}
	\label{eq6}
	\bar{E}_z(r,z)={\rm sign}(z)2\pi\sigma \left\{ \frac{2}{r_b^2-r_a^2}\sum_n e^{-\mu_n|z|}\left[\intop_{r_a}^{r_b}F_n(r)rdr\right]^2\Biggm/\intop_{R_{in}}^{R_{ext}}F_n^2(r)rdr\right\},
\end{equation}
Equation~(\ref{eq6}) is valid for all three configurations of the cloud: (i) annular cloud between the tube and the central electrode, (ii) angular cloud in the tube and no central rod, and (iii) solid cloud of radius $r_b$ in the cylindrical tube. The only difference between these configurations  is the explicit form of the eigenfunctions and corresponding eigenvalues.

\subsubsection{Annular Electron Cloud in the Presence of a Central Rod}

As a set of the eigenfunction, one can choose 
\begin{equation}
	\label{eq7}
	F_n(r)=N_0(\nu_n \rho_{in})J_0(\nu_n \rho)-J_0(\nu_n \rho_{in})N_0(\nu_n \rho),
\end{equation}
where $\nu_n\equiv \mu_n R_{ext}$ are the roots of the equation
\begin{equation}
	\label{eq8}
	N_0(\nu_n \rho_{in})J_0(\nu_n)-J_0(\nu_n \rho_{in})N_0(\nu_n)=0.
\end{equation}
Here $J_0(x)$ and $N_0(x)$ are zeroth order Bessel and Neumann functions, $\rho=r/R_{ext}$, and $\rho_{in}=R_{in}/R_{ext}$. 
The integrals in the numerator and denominator of the expression in Eq.~(\ref{eq6}) can be evaluated as
\begin{eqnarray}
	\label{eq10}
	\intop_{r_a}^{r_b}F_n(r)rdr=\left\{N_0(\nu_n\rho_{in})\left[\rho_bJ_1(\nu_n\rho_b)-\rho_aJ_1(\nu_n\rho_a)\right]\right.\nonumber\\
	-\left.J_0(\nu_n\rho_{in})\left[\rho_bN_1(\nu_n\rho_b)-\rho_aN_1(\nu_n\rho_a)\right]   \right\}R_{ext}^2/\nu_n\equiv g_n R_{ext}^2/\nu_n,
\end{eqnarray}
\begin{equation}
	\label{eq13}
	\intop_{R_{in}}^{R_{ext}}F_n^2(r)rdr=\frac{2R_{ext}^2}{\pi^2\nu_n^2}\left[\frac{J_0^2(\nu_n\rho_{in})}{J_0^2(\nu_n)}-1)\right]\equiv \frac{2R_{ext}^2}{\pi^2\nu_n^2}f_n.
\end{equation}
Here $\rho_{a,b}=r_{a,b}/R_{ext}$.
Finally, the expression for the axial component of the electric field $\bar{E}_z(z^\prime,z^{\prime\prime})$ produced in the cross section $z=z^\prime$ by the charged ring located at $z=z^{\prime\prime}$ can be written as 
\begin{equation}
	\label{eq14}
	\bar{E}_z(z^\prime,z^{\prime\prime})=2\pi\sigma {\rm sign}(z^\prime-z^{\prime\prime})G(|z^\prime-z^{\prime\prime}|),
\end{equation} 
where 
\begin{equation}\label{eq15}
	G(z)=\frac{\pi^2}{\rho_b^2-\rho_a^2}\sum_ne^{-\nu_n|z|/R_{ext}}g_n^2/f_n
\end{equation}

\subsubsection{Annular Electron Cloud --  No Central Rod}

For this configuration the eigenfunctions are $F_n(r)=J_0(\nu_n r/R_{ext})$, where $\nu_n$ are the zeros of the Bessel function $J_0(x)$. The electric field $\bar{E}_z(z^\prime,z^{\prime\prime})$ can be written in the same form as in  Eq.~(\ref{eq14}) but with different functions $G$, $g$ and $f$, given by
\begin{equation}
	\label{eq16}
	G(z)=\frac{4}{\rho_b^2-\rho_a^2}\sum_ne^{-\nu_n|z|/R_{ext}}\frac{g_n^2}{\nu_n^2f_n},
\end{equation} 
where 
\begin{equation}
	\label{eq17}
	g_n=\rho_bJ_1(\nu_n\rho_b)-\rho_aJ_1(\nu_n\rho_a)\hspace{3mm} {\rm and}\hspace{3mm} f_n=J_1^2(\nu_n).
\end{equation}

\subsubsection{Solid Cylindrical Electron Cloud}

The expression for the electric field of a solid electron cloud of radius $r_b$ ($r_a=0$) follows from Eq.~(\ref{eq17}):
\begin{equation}
	\label{eq18}
	G(z)=4\sum_ne^{-\nu_n|z|/R_{ext}}\frac{g_n^2}{\nu_n^2f_n},
\end{equation} 
where 
\begin{equation}
	\label{eq19}
	g_n=J_1(\nu_n\rho_b)\hspace{3mm} {\rm and}\hspace{3mm} f_n=J_1^2(\nu_n).
\end{equation}

\subsubsection{Approximation by Exponent}

It was shown in \cite{Tien, Rowe, Morey} that the function $G(z)$ for a solid cylindrical electron cloud can be approximated by the exponential
\begin{equation}
	\label{eq20}
	G(z)\simeq \exp(-|z|/\ell_C),
\end{equation}
where the characteristic scale of the Coulomb field  $\ell_C\simeq r_b/2$ and its value is practically independent of the ratio $r_b/R_{ext}=\rho_b$ for $\rho_b\leq 0.3$. 
Annular clouds, either in the presence of a central rod or not, are characterized by additional of geometrical parameters. Nevertheless, the exponential approximation Eq.~(\ref{eq20}) is still applicable.  
Note, that this is valid far from the potential well's edges. The Coulomb field of the charged plane slice is not symmetrical with respect to the slice position when the distance between the boundary electrodes and the slice is compared with $\ell_C$. This asymmetry can be properly considered, but its role is small when the distance between these electrodes is large as compared with $\ell_C$.

\subsection{Equation of Motion and the Steady-State Electron Density}

As mentioned above, the electron cloud can be represented as a set of charged rigid rings/disks with surface charge density $\sigma=en_edz$. The equation of motion of the rings/disks in the external electric field and the self-consistent Coulomb field of the electron space charge can be written as \cite{Bliokh-2021}
\begin{equation}
	\label{eq21}
	\frac{d^2z}{dt^2}=-\frac{e}{m}\frac{\partial\varphi}{\partial z}+2\pi\frac{e^2}{m}\intop_0^Ldz^\prime n_e(z^\prime){\rm sign}(z-z^\prime) G(|z-z^\prime|).
\end{equation}   
Here $\varphi$ is the external potential and $L$ is the system length. 
Assuming, that the characteristic spatial scale of the variation  of the external potential $\varphi$ and the electron density $n_e$ is large  compared to $\ell_C$, one can evaluate the integral in Eq.~(\ref{eq21}) as
\begin{equation}
	\label{eq22}
	\intop_0^Ldz^\prime n_e(z^\prime){\rm sign}(z-z^\prime) G(|z-z^\prime|)\simeq\intop_0^Ldz^\prime n_e(z^\prime){\rm sign}(z-z^\prime) e^{(|z-z^\prime|)/\ell_C}\simeq 2\frac{dn_e(z)}{dz}\ell_C^2.
\end{equation} 
It follows from Eq.~(\ref{eq22}) that the steady-state solution exists for 
\begin{equation}
	\label{eq23}
	n_e(z)=\frac{\varphi(z)}{4\pi e\ell_C^2}\simeq 5.7\cdot 10^8 \frac{\varphi(z)[kV]}{\ell_C^2[cm]} [cm^{-3}].
\end{equation}
Note, that  $\varphi(z)$ is the potential profile along the electron propagation path, i.e.,  $\varphi(z)=\varphi(z,r)_{r={\rm const}}$.


The steady-state density of a solid circular electron cloud of radius $r_d$ is
\begin{equation}
	\label{eq24}
	n_e(z)\simeq 2.3\cdot 10^9 \frac{\varphi(z)[kV]}{r_d^2[cm]} [cm^{-3}].
\end{equation}

\begin{figure}[tbh]
	\centering \scalebox{0.4}{\includegraphics{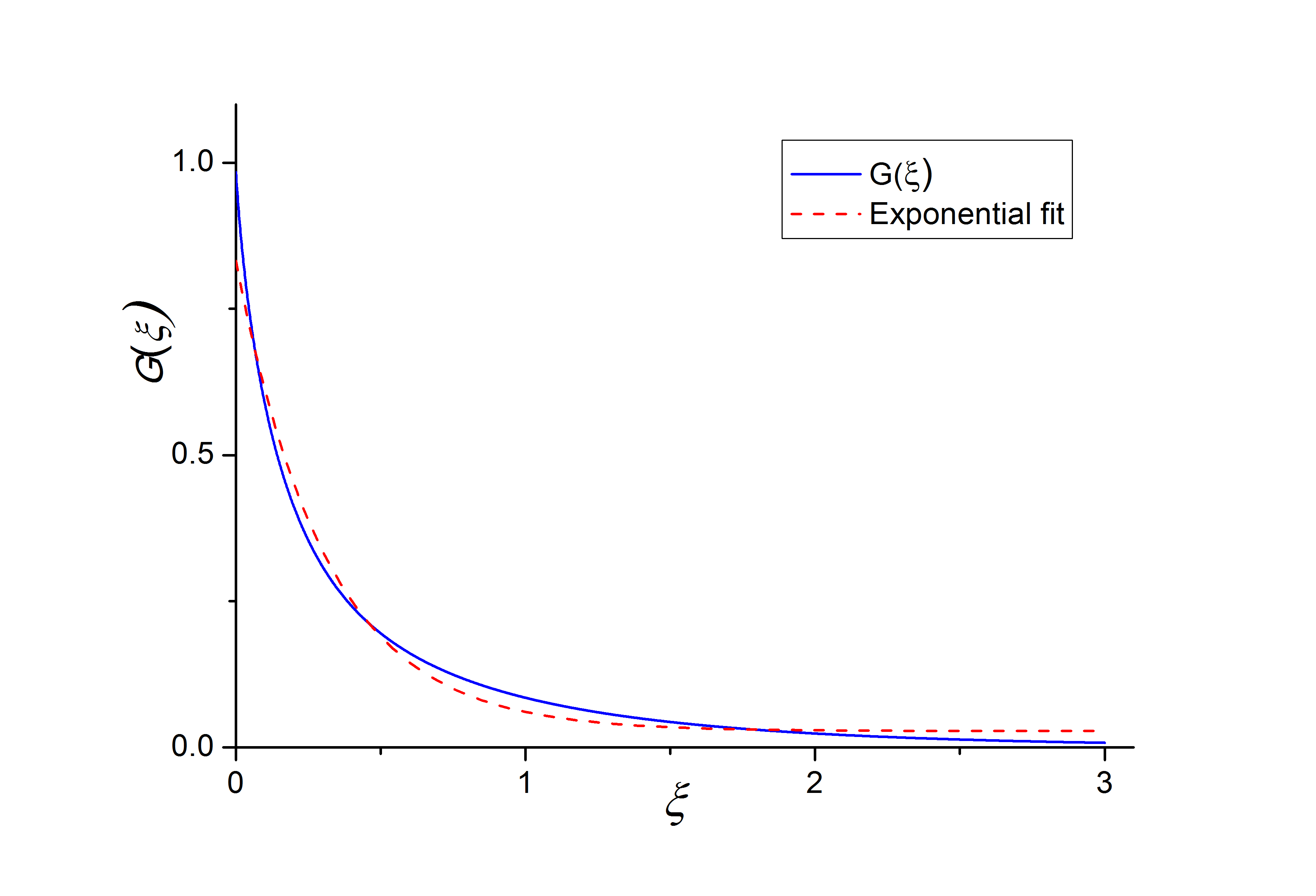}}
	\caption{Annular electron cloud including the central electrode. Exponential fit of function $G(\xi)$,  $\xi=z/\sqrt{r_b^2-r_a^2}$.  }
	\label{FigA1}
\end{figure}

The dimensions of the conductors and the electron cloud in the experiment described in Section III, are $r_a=0.7$cm, $r_b=0.9$cm, $R_{in}=0.3$cm, and $R_{ext}=2.0$ cm. The function $G$, defined by Eq.~(\ref{eq15}), and its exponential fit are shown in Fig. \ref{FigA1} as functions of the dimensionless distance $\xi=z/\sqrt{r_b^2-r_a^2}$. The characteristic scale of Coulomb field for these parameters is $\ell_C\simeq 0.56$cm. Note that $\ell_C$ remains the same when no central electrode is present.

\newpage
\newpage


\end{document}